\documentclass[
  journal=largetwo,
  manuscript=article-type,
  year=2020,
  volume=37,
]{cup-journal}

\usepackage{amsmath}
\usepackage[nopatch]{microtype}
\usepackage{booktabs}

\usepackage{graphicx}
\usepackage{siunitx}
\usepackage{amsmath}
\usepackage{mathtools}

\usepackage{float}
\usepackage{tabularx}
\usepackage{amsmath}
\usepackage{svg}

\title{Verifying the Australian MWA EoR pipeline I: \twocm sky model and correlated measurement density}

\author{J.~L.~B.~Line}
\affiliation{International Centre for Radio Astronomy Research, Curtin University, Perth, WA 6102, Australia}
\alsoaffiliation{ARC Centre of Excellence for All Sky Astrophysics in 3 Dimensions (ASTRO-3D)}
\email[J.~L.~B.~Line]{jack.l.b.line@gmail.com}

\author{C.~M.~Trott}
\affiliation{International Centre for Radio Astronomy Research, Curtin University, Perth, WA 6102, Australia}
\alsoaffiliation{ARC Centre of Excellence for All Sky Astrophysics in 3 Dimensions (ASTRO-3D)}

\author{J.~H.~Cook}
\affiliation{International Centre for Radio Astronomy Research, Curtin University, Perth, WA 6102, Australia}
\alsoaffiliation{ARC Centre of Excellence for All Sky Astrophysics in 3 Dimensions (ASTRO-3D)}

\author{B.~Greig}
\affiliation{School of Physics, University of Melbourne, Parkville, VIC 3010, Australia}
\alsoaffiliation{ARC Centre of Excellence for All Sky Astrophysics in 3 Dimensions (ASTRO-3D)}

\author{N.~Barry}
\affiliation{International Centre for Radio Astronomy Research, Curtin University, Perth, WA 6102, Australia}
\alsoaffiliation{ARC Centre of Excellence for All Sky Astrophysics in 3 Dimensions (ASTRO-3D)}

\author{C.~H.~Jordan}
\affiliation{International Centre for Radio Astronomy Research, Curtin University, Perth, WA 6102, Australia}
\alsoaffiliation{ARC Centre of Excellence for All Sky Astrophysics in 3 Dimensions (ASTRO-3D)}

\keywords{Astronomy data analysis, Reionisation, GPU computing} %% First letter not capped

\usepackage{xspace}

\newcommand{\RTS}{\texttt{RTS}\xspace}
\newcommand{\hyperdrive}{\texttt{hyperdrive}\xspace}
\newcommand{\Birli}{\texttt{Birli}\xspace}
\newcommand{\CHIPS}{\texttt{CHIPS}\xspace}
\newcommand{\OSIRIS}{\texttt{OSIRIS}\xspace}
\newcommand{\WODEN}{\texttt{WODEN}\xspace}
\newcommand{\pipe}{\texttt{AusEoRPipe}\xspace}

\newcommand{\twocm}{21-cm\xspace}

\usepackage{aas-macros}
\usepackage{hyperref} 
\hypersetup{colorlinks,citecolor=blue,linkcolor=blue,urlcolor=blue}

\begin{document}

\begin{abstract}
We present the first of two papers dedicated to verifying the Australian Epoch of Reionisation pipeline (\pipe) through simulation. The \pipe aims to disentangle \twocm radiation emitted by gas surrounding the very first stars from contaminating foreground astrophysical sources, and has been in development for close to a decade. In this paper, we build an accurate \twocm sky model that can be used by the \WODEN simulation software to create visibilities containing a predictable \twocm signal. We verify that the power spectrum estimator \CHIPS can recover this signal in the absence of foregrounds. We also investigate how measurements in Fourier-space are correlated, and how their gridded density affects the power spectrum. We measure and fit for this effect using Gaussian-noise simulations of the MWA phase I layout. We find a gridding density correction factor of 2.651 appropriate for integrations equal to or greater than 30 minutes of data, which contain observations with multiple primary beam pointings and LSTs. Paper II of this series will use the results of this paper to test the \pipe in the presence of foregrounds and instrumental effects.
\end{abstract}

\section{Introduction}
\label{sec:intro}
Constraining when and how the first stars were formed has been a goal of astronomers for decades. By measuring cosmically-redshifted \twocm radiation from the hydrogen gas surrounding those first stars, one can conceivably map the effects of the ionising radiation coming from them, to infer their properties as a function of redshift. Unfortunately, the detection of \twocm radiation from the Epoch of Reionisation (EoR) (redshifted to the $~$50-250$\,$MHz frequency range), is impeded by a myriad of astrophysical sources, including Active Galactic Nuclei, Radio Galaxies, Supernova Remnants, and the diffuse synchrotron radiation emitted from the Milky Way. These foregrounds drown out the \twocm signal~\citep{Furlanetto2006b}, and must either be subtracted from the data or avoided. One approach that somewhat naturally separates foregrounds from the signal is via measuring the power spectrum (PS). The PS can be used to garner spatial information of the effects of reionisation by the first stars, and infer their properties. Furthermore, Fourier transforming over frequency to obtain a PS results in the foregrounds and signal manifesting in different areas of the PS given their differing spectral behaviours. Recent upper limits on the EoR have come from: Hydrogen Epoch of Reionisation Array~\citep[HERA,][]{DeBoer2016}; Low-Frequency Array~\citep[LOFAR,][]{VanHaarlem2013}; The Murchison Widefield Array~\citep[MWA,][]{Tingay2013,Wayth2018}; New Extension in Nançay Upgrading LOFAR~\citep[NenuFar,][]{Zarka2012}.

There are a number of challenges which must be overcome to make the statistical detection of the \twocm signal. As well as nullifying the aforementioned foregrounds~\citep[e.g.][]{Cook2022,Acharya2023}, one must deal with ionospheric refraction~\citep[e.g.]{Elder2021,Chege2022}, terrestrial interference~\citep[e.g.][]{Offringa2019,Wilensky2023}, and a myriad of instrumental calibration effects~\citep[e.g.][]{Kern2020, Chokshi2021, Mevius2022, Kolopanis2023}, all of which cause limiting systematics. These challenges have forced a decade of not only hardware but also software development, to calibrate and treat the data with extreme precision to uncover the underlying signal. Recent best-effort PS limits include~\citet{Mertens2020,Trott2020,Rahimi2021,HERA2023,Munshi2024}.

Given that measuring the \twocm PS is as much a software as it is an observational experiment, great care must be taken to understand what biases and systematics the software pipeline imparts upon the data. Given the limiting systematics listed above, a natural route is via simulation, where one can control which systematics are injected into a data set, with a ground ``truth'' to be recovered. In short, if you inject a known \twocm signal into a data set, process through your pipeline and recover what you injected, you gain confidence in your pipeline. This approach has been taken by a number of authors, including~\citet{Barry2019_pipeline} using MWA data and~\citet{HERA_verify} with HERA data. To date, the Australian EoR pipeline (\pipe), which is designed to process MWA data, has not been tested from end to end. This is the goal of this series of papers. This paper aims to test the gridding and PS estimation aspects of the pipeline; the following paper in this series will focus on calibration.

The native output of an interferometer is a visibility, a sampling of the Fourier transform of the visible sky. There are two main ways to simulate a \twocm signal in visibility space. One can either make an analytic model that produces similar statistics in a PS, to directly generate in visibility space, or can create an image based model and derive visibilities from that. The former is far less computationally expensive, but the latter locks in projection and geometrical effects inherent to true interferometric observations. Having an image-based \twocm model locked to celestial coordinates allows for simulated observations at various hour angles, to probe the effects of changing instrumental primary beam patterns and visibility sampling.

Simulating an image-based \twocm signal for the MWA is non-trivial however, given the sky coverage of the instrument. The full-width half-maximum of the main lobe of the primary beam of the MWA can be greater than $20 \times 20$ square degrees, and $50 \times 50$ square degrees down to the 1\% beam level. Simulating a realistic EoR volume capable of covering this sky-area is computationally demanding. Fortunately, \citet{Greig2022} generated a simulated EoR volume large enough, which we use in this work and detail in Section~\ref{sec:21cm_skymodel}. Given the resolution of the instrument is of order an arcminute, the number of pixels required to represent \twocm lightcones from this EoR volume is $\mathcal{O}(10^7)$. Down-sampling this model can affect the resultant statistics of the signal. For this reason, we use the \WODEN simulation package~\citep{Line2022}, capable of ingesting sky models with millions of components, and producing realistic simulated MWA visibilities. As \WODEN is GPU-accelerated, these simulations can be run in a reasonable time frame (see Section~\ref{subsec:grid_dense_correct} for details). Importantly, \WODEN is also capable of simulating both the \twocm signal and the foregrounds at the expected power levels, which can be many orders of magnitude apart. This allows us to test signal recovery in the presence of foregrounds without boosting the amplitude of the \twocm signal.

This paper is organised as follows. In Section~\ref{sec:pipeover} we briefly overview the \pipe and outline what simulated data are necessary to test them. In Section~\ref{sec:21cm_skymodel} we detail the construction of a \twocm sky model compatible with \WODEN. In Section~\ref{sec:gauss_noise} we verify the steps taken to project the \twocm model into celestial coordinates, by applying them to a sky model of Gaussian noise. We also investigate the effects of gridding visibilities when generating power spectra, and fit for the effect this has on estimating the resultant power. In Section~\ref{sec:recover_21cm} we test the \twocm model and how well the \pipe can recover the expected PS. We discuss our results in Section~\ref{sec:discuss}, and conclude in Section~\ref{sec:conclusion}.

\section{Pipeline overview}
\label{sec:pipeover}
The \pipe is designed to take raw data from the MWA correlator, calibrate for instrumental and ionospheric effects, subtract foregrounds, and then estimate a PS. Here we list the different software packages that make up the \pipe, and suggest simulations needed to test them. This Section is intended to give the reader a broad overview of the \pipe and the overall goals of this series of papers, to give this paper context. At the end of this Section, we detail which parts of the work are covered in this paper.

\Birli\footnote{\url{https://github.com/MWATelescope/Birli}} is designed to take raw MWA data and preprocess it, including geometric correction, averaging, and radio frequency interference (RFI) flagging (via \texttt{AOFlagger}~\citep{Offringa2010} or \texttt{SINSS}~\citep{Wilensky2019}). \Birli has been written to replace \texttt{Cotter}~\citep{Offringa2015}, to have one unifying preprocessing package that works with both legacy and new MWA data~\citep[see][for details of the old and new MWA correlators]{Morrison2023}. Given \Birli has been tested to reproduce results out of \texttt{Cotter}, which has been tested and used within a number of MWA pipelines, we choose not to test \Birli via simulation. To do so however, one would need to simulate raw MWA data, which for the original correlator would mean producing visibilities with no phase tracking, adding a frequency dependent bandpass from the MWA polyphase band filter, and add in realistic RFI.

\hyperdrive\footnote{\url{https://github.com/MWATelescope/mwa_hyperdrive}}~(Jordan et al. \textit{submitted}), takes either raw or preprocessed data, calibrates it, and then subtracts foregrounds. Built to replace the \RTS~\citep{Mitchell2008}, \hyperdrive also creates model visibilities to calibrate against via an image-based sky model. It initially performs a direction independent (DI) calibration step, that derives a gain for each MWA receiving element (tile). All frequencies are calibrated independently. After DI calibration, foregrounds can be subtracted. The Fourier transform along frequency to create the PS is expected to separate the \twocm signal from the foregrounds, as foreground emission is supposed to be spectrally smooth, whereas the \twocm signal has spectral structure. Incomplete sampling of the $u,v$ plane however causes mode mixing between the two, as well as other frequency-related instrumental effects. A detection therefore relies on the foregrounds being subtracted without injecting false spectral structure, as to preserve the underlying \twocm signal. To test this functionality, a simulation needs to contain both a \twocm signal and foregrounds, as well as frequency dependent instrumental gain errors. One can also add frequency-dependent effects like cable reflections~\citep[e.g.][]{Ewall-Wice2016} and the bandpass to investigate how calibrating each frequency channel independently performs. These effects are direction-independent and so can be added to visibilities post simulation (functionality which is in development in \WODEN). Further simulations including RFI and ionospheric refraction would test how robust the calibration is to environmental influences. These effects are direction dependent and so must be adding during the calculations of the visibilities, which \WODEN is currently incapable of.

\CHIPS~\citep{Trott2016} takes the calibrated and subtracted visibilities and grids them into a spectral cube using an optimised Blackman-harris kernel. It then Fourier-transforms along frequency, and then either cylindrically averages to create a two dimensional PS (2D PS) or spherically averages to create a one dimensional PS (1D PS). Integrations of tens to hundreds of hours of data into a 1D PS have set the limits that have been released by the \pipe to date. The minimal test for \CHIPS is to simulate a set of visibilities with a known \twocm signal, and see if \CHIPS can recover it. Beyond that, taking in calibrated visibilities out of \hyperdrive, derived from simulations with realistic instrumental effects and foregrounds, allows us to probe the effects of calibration on the resultant PS.

\WODEN\footnote{\url{https://github.com/JLBLine/WODEN}}, as previously mentioned, simulates MWA visibilities from a sky based model. It can use the same primary beam model as \hyperdrive, the FEE beam~\citep{Sokolowski2017} (via the \texttt{mwa\_{}hyperbeam}\footnote{\url{https://github.com/MWATelescope/mwa_hyperbeam}} package), and creates visibilities that can be fed directly into either \hyperdrive or \CHIPS. \WODEN calculates the interferometric measurement equation for every component in the sky model. Each component can either be a point source (a dirac-delta function upon the sky), an elliptical Gaussian, or a Shapelet model~\citep[see][for more detail on Shapelets]{Line2020}. The measurement equation encodes baseline and sky projection effects inherently, making the simulator accurate across the large field of view of the MWA. The caveat being that the sky model must be broken into the small components listed here, meaning representations of large-scale structure such as the galactic diffuse emission need millions of components. Aside from the primary beam model, the \WODEN code base is entirely independent to \hyperdrive. Although they use the same methodology to generate visibilities, this redundancy in code is by design; any internal bug that may cancel out when generating visibilities and calibrating them all with the same code base is avoided.

Along with the aforementioned packages, extensive quality metrics are generated at various stages (Nunhokee et al. \textit{submitted}). These metrics are used to cull any observations that are deemed unusable due to insurmountable instrumental effects, ionospheric conditions, RFI events, contamination by bright sources in the primary beam sidelobes, and other effects. The simulations detailed above could be used to check whether these metrics are able to catch poor data, and check whether the limits set for each metric still allow for a detection of the \twocm signal.

In this paper, we ignore calibration entirely, and focus on generating a realistic \twocm sky model to simulate through \WODEN and directly input into \CHIPS. As visibilities are additive, once we are able to generate accurate \twocm simulations, we can add a variety of foregrounds and instrumental effects to test the calibration and subtraction step, without the need to rerun the more expensive \twocm simulations. We leave testing \hyperdrive to the second paper in this series.

\subsection{MWA EoR observing}

The MWA is a low-frequency radio interferometer, with receiving elements consisting of 4$\times$4 grids of dual linear-polarisation bow-tie antennas. These `tiles' are electronically steered through beamforming. To estimate the \twocm PS, the MWA EoR collaboration have identified a number of fields with lower sky temperatures, which have now been observed for hundreds of hours~\citep[e.g.][]{Trott2020}, in an attempt to average over thermal noise. Given the electronic beamforming, this has resulted in a drift-and-shift observational campaign, where the target field drifts through a number of pointings, with scheduling keeping the field centre as close to the primary beam centre as possible~\citep[see][for more details]{Jacobs2016}. 

In this paper we focus on the EoR0 field, centred at RA, DEC $=0^h, -27^\circ$, observed between frequencies of 167-198$\,$MHz (known as the high-band). These parameters have consistently yielded the best MWA limits~\citep[e.g.][]{Barry2019,Trott2020,Chege2022}. We also focus on five pointings, labelled -2 through +2~\citep[c.f.][]{Beardsley2016}, where -2 means two pointings before the zenith 0 pointing, and +2 meaning two pointings after 0 pointing. There is an approximate 15$^\circ$ shift between pointings, all along the meridian. Examples are shown in Figure~\ref{fig:pointings}. MWA EoR data are typically taken in two minute chunks known as an observation.

\section{\twocm sky-model}
\label{sec:21cm_skymodel}

We use the \twocm lightcone detailed in~\citet{Greig2022}, which is derived from a simulated EoR cube with sides of length $7.5\,$Gpc. This cube was generated using a simplified version of \texttt{21CMFACST}~\citep{Mesinger2007,Mesinger2011}, and was spectrally sampled at 80$\,$kHz to match typical MWA EoR analysis parameters. This methodology generates a volume that projects to a sky coverage of $\sim 50 \times 50$ square degrees, which is essential given the footprint of the MWA primary beam. A number of steps are necessary to translate the lightcone box into a \WODEN sky model. The box is a collection of 2D projected 21\,cm intensity distributions, each at a unique redshift. These cartesian projections are taken by slicing a full 3D simulation volume at regular redshift intervals, as the simulation is evolved with time. This leaves us with a number of transformations needed to allow \WODEN to ingest the model:

\begin{itemize}
  \item Each 2D \twocm slice has a pixel resolution $\Delta x$ in cMpc; each slice therefore has a different angular resolution, as this quantity is redshift dependent. To run efficiently, \WODEN needs a grid of pixels constantly sampled in RA/DEC, necessitating a redshift dependent angular interpolation. 
  \item The box is regularly sampled in redshift ($\Delta z$), which results in irregular sampling in frequency ($\Delta \nu$). Interferometric data is sampled regularly in frequency, necessitating a second interpolation over frequency.
  \item The box is in units of mK. \WODEN ingests units of Jansky (Jy), which is an integrated flux density. The conversion from mK to Jy/sterrad is straightforward, however, the physical volume of each pixel changes as a function of frequency. Along with the angular extent, the pixel volume is determined by $\Delta z$. Given the interpolation over angle and frequency, this volume change must be taken into account, as it effects the resultant variance of the \twocm map. This effect is noted in the script \texttt{make\_{}flat\_{}spectrum\_{}eor.py}\footnote{\url{https://github.com/RadioAstronomySoftwareGroup/pyradiosky/blob/main/scripts/make_flat_spectrum_eor.py}} in the Python package \texttt{pyradiosky}.
\end{itemize}

In the following, we use the cosmology assumed by~\citet{Greig2022}, a $\Lambda$CDM cosmology with: $H_0=68; \Omega_{M}=0.31; \Omega_{\Lambda}=0.69; \Omega_{b}=0.048$. We use \texttt{astropy}~\citep{astropy:2013, astropy:2018, astropy:2022} for all cosmological calculations.

We choose to interpolate all slices to the finest angular resolution in the box to retain as much angular structure as possible. We call this the reference redshift, $z_{\mathrm{ref}} = 7.5693$. Given $\Delta \theta = \Delta x / \mathcal{D}(z)$, where $\mathcal{D}$ is the comoving distance, we interpolate to an angular resolution of $\sim27$ arcseconds. We found interpolating the Cartesian slice directly into a \texttt{TAN FITS} projection~\citep[a gnomic projection; see][for details]{FITS2002} centred at RA, DEC $=0^h, -27^\circ$ returned the expected PS. We experimented with bilinear and bicubic interpolation, but found significant signal loss in the resultant PS. It's possible some variant of Gaussian Process Regression may be more effective, however for the purposes of this work we found a simple nearest-neighbour approximation was sufficient. Similarly, we applied a nearest-neighbour interpolation along the frequency axis, as again we saw signal loss in the final PS when applying linear or cubic interpolation. Along lines of sight, the \twocm signal rapidly fluctuates between positive and zero, and so these kinds of interpolation tend to create false signal. Once these interpolations have been applied, we then scale for the change in volume of the pixel ($\mathrm{V}_{\mathrm{pix}}(z)$) and its effect on the variance. We do this for each redshift by calculating via the differential comoving volume and multiplying by $\Delta z$. We then scale each redshift slice by a factor $C_{\mathrm{pix}}$ given by
\begin{equation}
    C_{\mathrm{pix}} = \sqrt{\frac{\mathrm{V}_{\mathrm{pix}}(z)}{\mathrm{V}_{\mathrm{pix}}(z_{\mathrm{ref}})}},
\end{equation}
with the square root ensuring the variance is scaled by the pixel volume.

\begin{figure}[h!]
    \centering
    \includegraphics[width=\columnwidth]{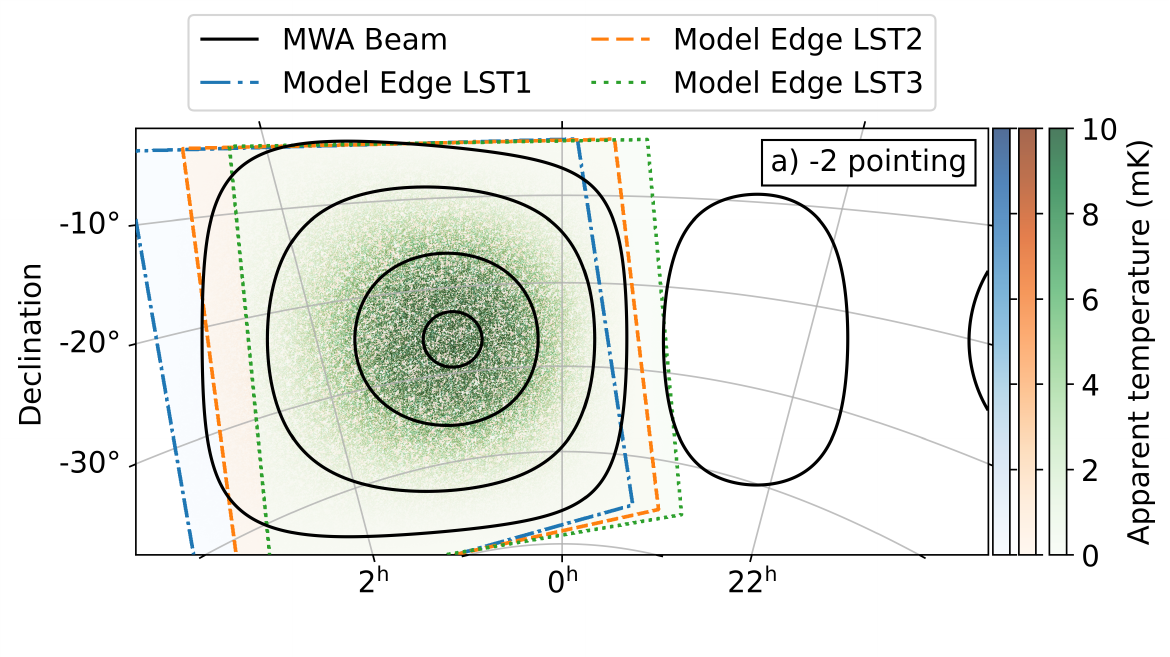}
    \includegraphics[width=\columnwidth]{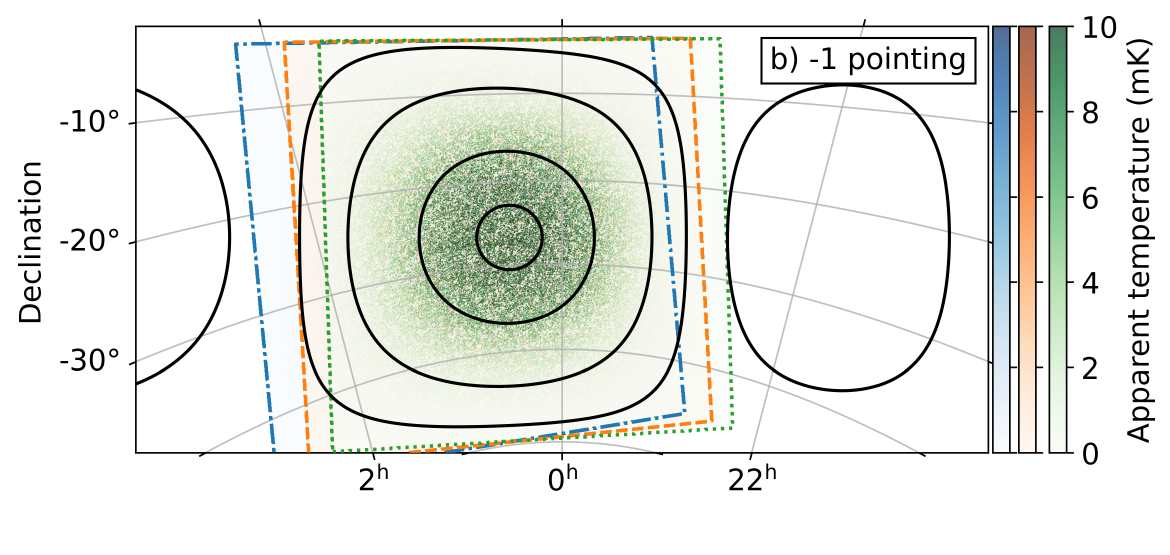}
    \includegraphics[width=\columnwidth]{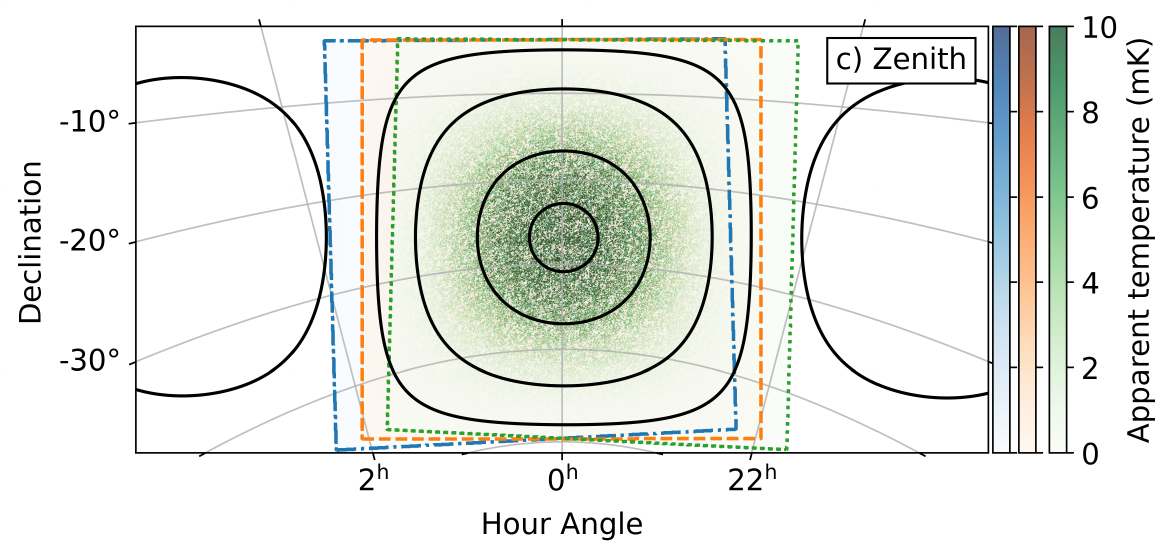}
    \caption{Demonstration of the MWA primary beam and its interaction with the \twocm sky model, via the: a) -2 pointing; b) -1 pointing; c) Zenith pointing. Each plot is locked to the observer in HA/Dec. The solid black lines represent the instrumental Stokes I primary beam pattern contoured at 1\%, 10\%, 50\%, and 90\% power levels. The coloured images show the apparent \twocm sky model after attenuation by the primary beam at three different LSTs, with the corresponding colour lines marking the edges of the full \twocm sky model. Both the beam and sky model are shown at 167$\,$MHz, where the primary beam is largest for a high-band observation. Note that the +1, +2, pointings as described in Section~\ref{sec:recover_21cm} are simply westward reflections of the -1, -2 pointings, so aren't reproduced here for brevity.}
    \label{fig:pointings}
\end{figure}

Once we have interpolated, scaled, and transformed each slice in Jy/sterrad, we convert each pixel into a point source with units of Jy by multiplying by the pixel solid angle.
Ideally, we would the find some way to tile these maps to give an all-sky \twocm sky model. However, given we cannot interpolate the box without signal loss, it is computationally infeasible. At the lowest frequencies, the number of point sources necessary would approach 200 million. However, given the extreme volume of the simulated \twocm box, without tiling, the sky model already covers the main lobe of the MWA primary beam down to the $~1$\% power level for pointings of interest (see Figure~\ref{fig:pointings}).

To investigate whether the projection steps detailed in this Section induce any form of signal loss or bias, we apply them to a purely Gaussian noise simulation in Section~\ref{sec:gauss_noise}. We report of the results of the \twocm sky model in Section~\ref{sec:recover_21cm}.

\section{White Gaussian noise sky simulation}
\label{sec:gauss_noise}

The power spectrum $P(k)$ of a White Gaussian noise distribution $\mathcal{N}$ of mean $\mu=0$ and variance $\sigma^2$ is proportionate to the variance, i.e. $P(k) \propto \sigma^2$. By creating a sky model of purely White Gaussian noise, we can therefore predict an output $P(k)$, and check the normalisation of the \pipe. Given we have outlined a method to project a \twocm box from units of cMpc in Section~\ref{sec:21cm_skymodel}, we can simply repeat the entire process, starting from a box of purely White Gaussian noise (from hereon in referred to as the Gaussian noise simulation). We choose to set $\sigma^2=16\textrm{mK}^2$ as this yields a power comparable to those expected from the \twocm signal. We typically report the power spectrum in units of mK$^2$ Mpc$^3$ $h^{-3}$, where $h$ is the reduced Hubble constant; we use $h=0.68$ in this work. We can therefore predict our expected noise power spectrum $P_N$ via
\begin{equation}
    P_N(k) = \sigma^2 \mathrm{V}_{\mathrm{pix}}(z_\mathrm{ref}) h^3 = 8.923 \, \mathrm{mK}^2 \mathrm{Mpc}^3 h^{-3}
\end{equation}

% \pagebreak

\begin{figure*}[h!]
  \centering
  \includegraphics[width=\textwidth]{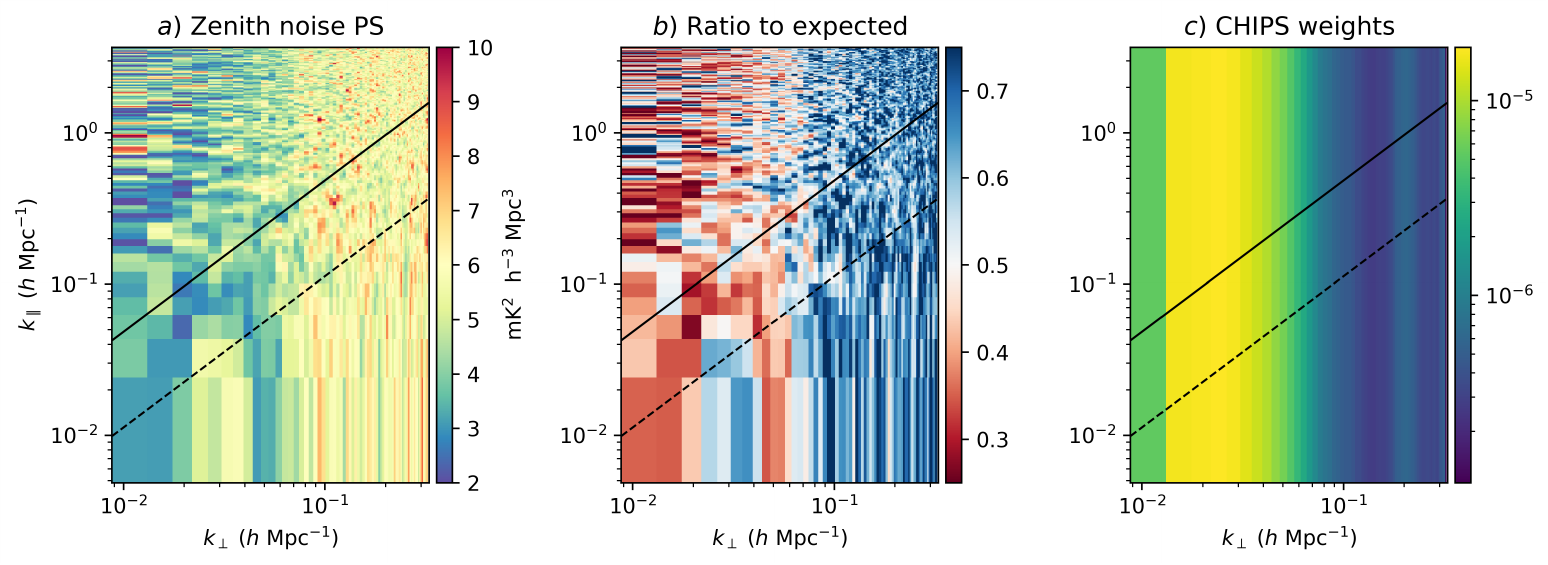}
  \caption{Data from a single zenith observation, without any correction for gridding density, where: \textit{a)} shows the 2D PS; \textit{b)} the ratio to the expected value; \textit{c)} the \CHIPS gridding weights. $k_\perp$ modes are derived from the instrument sampling in the visibility $u,v$ plane, averaged down to one dimension, and $k_\parallel$ from the fourier transform of the visibilities along frequency. The \CHIPS gridding weights therefore show the $u,v$ gridding density averaged into one dimension.}
  \label{fig:2D_noise_PS_nofit}
\end{figure*}

We use \WODEN\footnote{For this paper we use the docker v2.0 container of \WODEN, installed via \texttt{docker pull docker://jlbline/woden-2.0}} to simulate a single zenith pointing observation with an 8$\,$s time cadence and 80$\,$kHz frequency resolution, using a frequency-interpolated version of the FEE beam. We use the convention that Stokes I = (XX + YY) / 2, where XX and YY are the two linear polarisations. Given we are using simulated data with isotropic and unpolarised sky models, there is little difference between XX and YY, and therefore the power spectra shown are effectively Stokes I. All PS shown in this paper are from the north-south aligned polarisation. We run \CHIPS on the simulated visibilities, to produce the 2D PS as shown in Figure~\ref{fig:2D_noise_PS_nofit}. All \CHIPS plots were generated using the Python package \texttt{CHIPS\_wrappers}\footnote{\url{https://github.com/JLBLine/CHIPS_wrappers}}.

Figure~\ref{fig:2D_noise_PS_nofit} demonstrates a known effect of estimating the PS from gridded visibilities, where the power inferred is lower where the density of gridded visibilities is higher, as observed by~\citet[][see Appendix A]{Barry2019_pipeline}. Note we have not corrected for this in Figure~\ref{fig:2D_noise_PS_nofit}, which is normally done by multiplying by a factor two (as found by~\citet{Barry2019_pipeline}), to explicitly show the effect. 

Visibilities are gridding using a kernel which is designed to minimise aliasing and match the area of the MWA primary beam~\citep{Trott2016}. Given the array layout, and the need to match the kernel to the primary beam, the kernel has an unavoidable footprint large enough to overlap with neighbouring gridded visibilities. Depending on the covariance between neighbouring visibilities, and the density of the gridded visibilities, the power estimated will therefore vary as a function of $u,v$ gridding location. Using simulations, \citet{Barry2019_pipeline} found a factor of two correction was sufficient as a normalisation factor for modes of interest. This estimate was made using the \texttt{FHD}/\texttt{$\varepsilon$ppsilon} pipeline and a model of the MWA primary beam as a gridding kernel, whereas \CHIPS uses a Blackman-Harris kernel. This estimate was also made using 2$\,$s resolution data, rather than 8$\,$s. Given these differences, we investigate the gridding density correction factor in Section~\ref{subsec:grid_dense_correct} using \CHIPS.

\subsection{Gridding density correction factor}
\label{subsec:grid_dense_correct}
\begin{figure*}[h!]
  \centering
  \includegraphics[width=\textwidth]{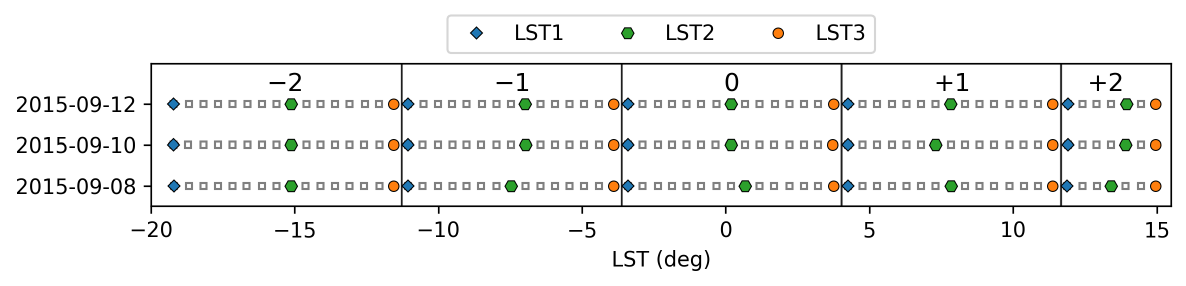}
  \caption{Observations simulated in this paper, based on two real nights of MWA EoR0 observing. The blue diamonds show the simulated LST1 subset,  green hexagons the LST2 subset, and orange circles the LST3 subset. Each hollow square shows a different two minute snapshot which was not simulated but exists in the real data set. Dividing lines and labels show the changes from pointings -2 through +2. Note observational constraints mean there are less +2 pointings.}
  \label{fig:obs_selection}
\end{figure*}

The gridding density is directly affected by the $u,v$ coverage, which in turn is dictated by the array layout and phase centre. The covariance of neighbouring $u,v$ points will therefore be affected by LST (as the phase centre is always set to EoR0 field centre). The primary beam pointing will also have an effect as this changes the amplitude and spectral behaviour of the visibilities. To thoroughly investigate the gridding density we therefore simulate a number of observations spanning realistic ranges of LST and pointings. To do so, we take the first (subset LST1), central (subset LST2), and final observation (subset LST3) across five different pointings (-2 to +2) from three real nights of MWA phase I EoR observing (see Figure~\ref{fig:obs_selection}). For the LST2 subset, we select the observation where the primary beam pointing is closest to the EoR0 field centre, to maximise the beam coverage over the sky model coverage. As the visibilities are normally averaged to $8\,$s, $80\,$kHz we simulate at this cadence to save compute time. Each observation takes a total of $\sim32\,$GPU hours. For 45 simulated observations this is a total of 60$\,$GPU days. We use the Pawsey Garrawarla cluster\footnote{\url{https://pawsey.org.au/systems/garrawarla/}}, and split each simulation across 24 GPUs, meaning these simulations take less than 3 days real-time.

\begin{figure*}[h!]
  \centering
  \includegraphics[width=0.8\textwidth]{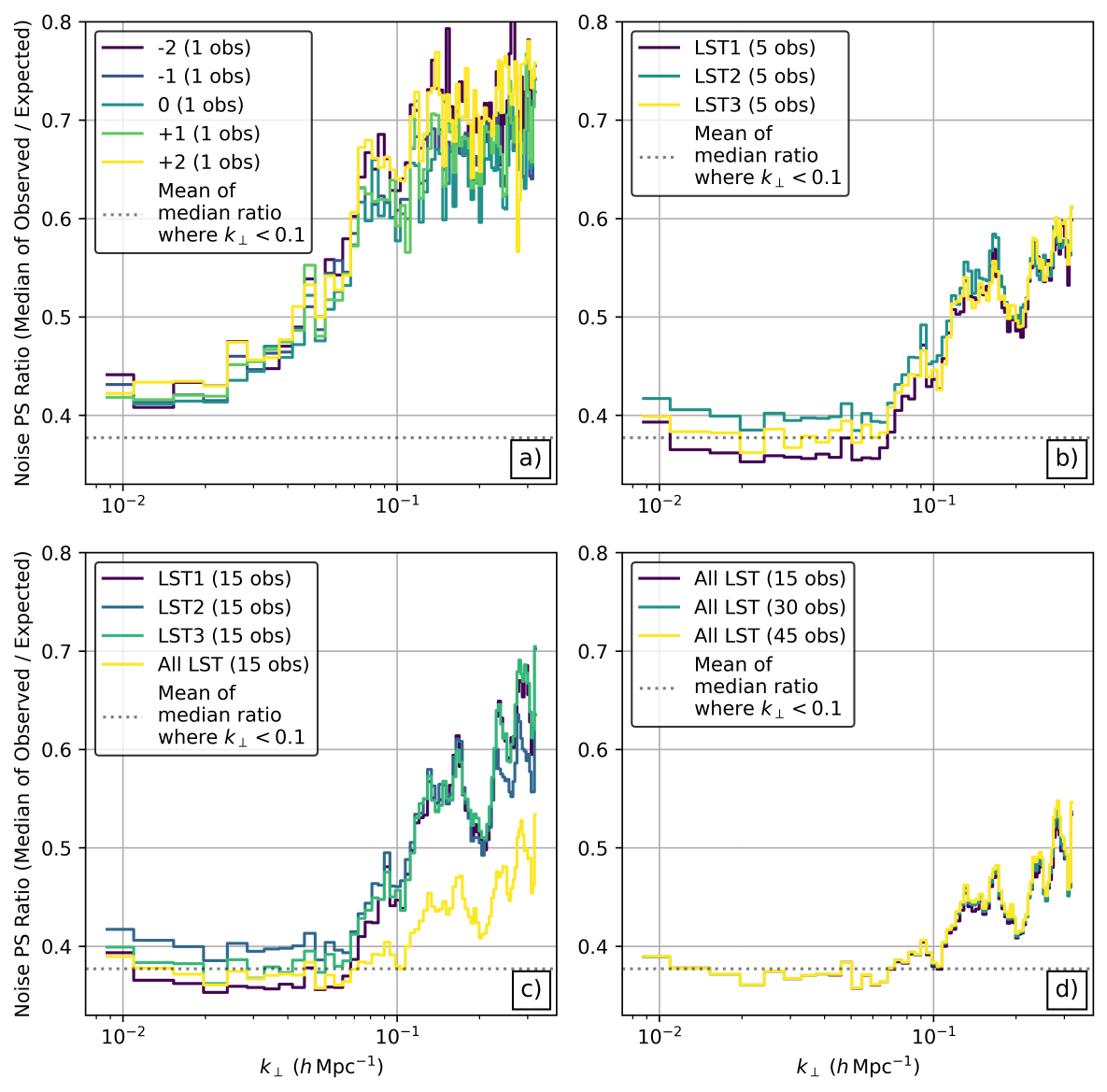}
  \caption{The median ratio of the recovered noise PS to expected value, as a function of $k_\perp$, for: \textit{a)} the five different pointings, each for a single observation; \textit{b)} the three different LST subsets integrated over 5 observations; \textit{c)} the three different LST subsets integrated over 15 observations; \textit{d)} all three LST subsets integrated together over 15, 30, and 45 observations. The horizontal dashed line shows the mean of the median ratios for $k_\perp < 0.1$ as measured from the bottom right plot. This dashed line Any dataset which is a multiple of five observations has an even split across the five pointings.}
  \label{fig:measure_gridding_factor}
\end{figure*}

By running \CHIPS on various combinations of these observations, we can investigate both primary beam pointing and LST effects. All integrations over multiple observations are done coherently, i.e. all observations are gridded to the same $u,v$ grid, which is then used to estimate the PS. We take the median of the ratio of the observed PS to the expected value as a function of $k_\perp$ (i.e. take the median along the y-axis of the middle panel in Figure~\ref{fig:2D_noise_PS_nofit}) to see the effects of gridding, as shown in Figure~\ref{fig:measure_gridding_factor}. We take the median, rather than the mean, as the distribution of ratios along each $k_\perp$ bin display significant skew.

Figure~\ref{fig:measure_gridding_factor}{\color{blue}a} shows that for a single observation, the estimated power is consistent across pointings, with lower power estimated a low $k_\perp$ (higher gridding density). This shows the beam volume correction applied internally in \CHIPS returns a consistent power level across the five pointings. Figure~\ref{fig:measure_gridding_factor}{\color{blue}b} shows that the power recovered from the three LST subsets over 5 observations is roughly consistent, but the power recovered does vary somewhat. This recovered power depends on the exact $u,v$ coverage, and the way the primary beam interacts with the sky model. Figure~\ref{fig:measure_gridding_factor}{\color{blue}c} shows that when integrating over 15 observations for each LST subset, the recovered power at low $k_\perp$ is consistent with the power recovered when integrating over 5 observations. However, the power recovered at higher $k_\perp$ is actually higher than at when integrating over 5 observations. This highlights the fact that the power estimated comes from a combination of not just the amount of data gridded, but the covariance between the neighbouring gridded visibilities. This is further highlighted by showing that integrating the three LST subsets together across 15 observations yields significantly lower power at high $k_\perp$. High $k_\perp$ is derived from longer baselines, where the drift of a baseline $u,v$ coordinate over time is greatest. The LST1 and LST3 groups are also closely separated in time, yielding close gridding locations, but overlap as the beam pointing changes. The amplitude and the spectral behaviour of the primary beam therefore sharply changes on the sky, reducing the covariance between neighbouring gridded visibilities. This results in the lower power recovered. Figure~\ref{fig:measure_gridding_factor}{\color{blue}d} shows that when integrating more and more data with the same combination of LST coverage and pointings, there is no change in power at low $k_\perp$.

\begin{figure}[h!]
  \centering
  \includegraphics[width=\columnwidth]{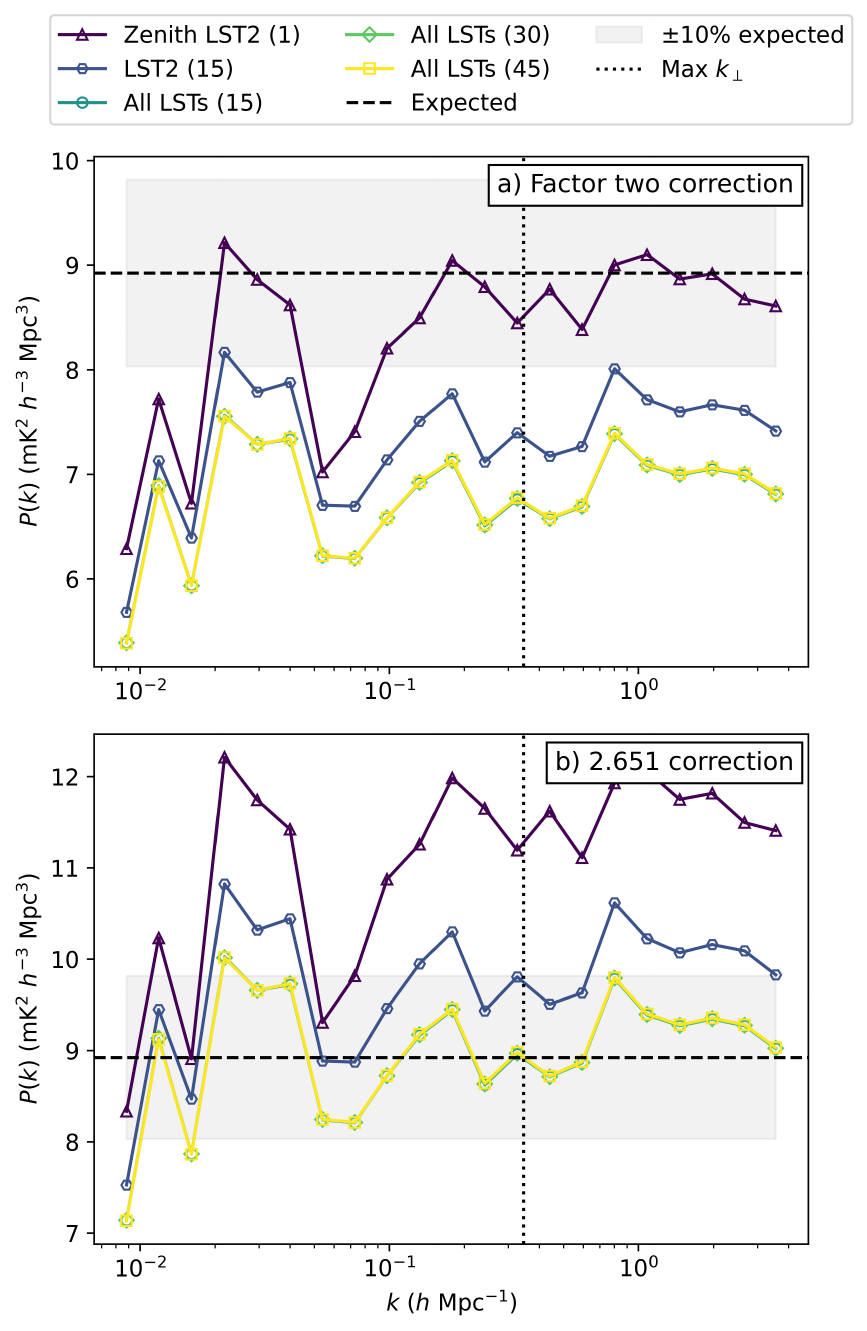}
  \caption{1D PS for various integrations of the Gaussian noise simulation, where \textit{a)} a factor of two was used to normalise for gridding density and \textit{b)} a factor of 2.5845 was used. The vertical dashed line shows the maximum $k_\perp$-mode.}
  \label{fig:compare_recovery_noise}
\end{figure}

Given the stability of the recovered power when combining all LST subsets with an integration of 15 observations or more, one could fit a functional form to derive a correction. We investigate this approach in~\ref{app:fit_gridding_factor}, and find little to no difference in using a fit when compared to just applying a scalar normalisation factor. PS upper limits are derived from integrating hours of data to overcome thermal noise, and modes where $k_\perp > 0.1\,h\,\textrm{Mpc}^{-1}$ are typically cut when using the \pipe. To obtain a normalisation factor for the gridding density we therefore simply take the mean of the median observed ratios for the integrations over all LSTs and pointings where $k_\perp < 0.1\,h\,\textrm{Mpc}^{-1}$. This value is plotted as the horizontal grey dashed line in Figure~\ref{fig:measure_gridding_factor}. Inverting this gives a normalisation factor of 2.651. We apply this normalisation and create a number of 1D PS for various integrations, shown in Figure~\ref{fig:compare_recovery_noise}, and compare the outcomes to the previous correction factor of two. This shows that the new normalisation factor causes too much power to be recovered for a single observation, and when only using the LST2 subset, but recovers the correct power to within 10\% for most $k$ modes when integrating over all LSTs and pointings. The factor two correction (coincidentally) does well for a single observation, but does not recover the correct power for higher integrations.

\section{Recovered \twocm signal}
\label{sec:recover_21cm}

\begin{figure*}[h!]
  \centering
  \includegraphics[width=\textwidth]{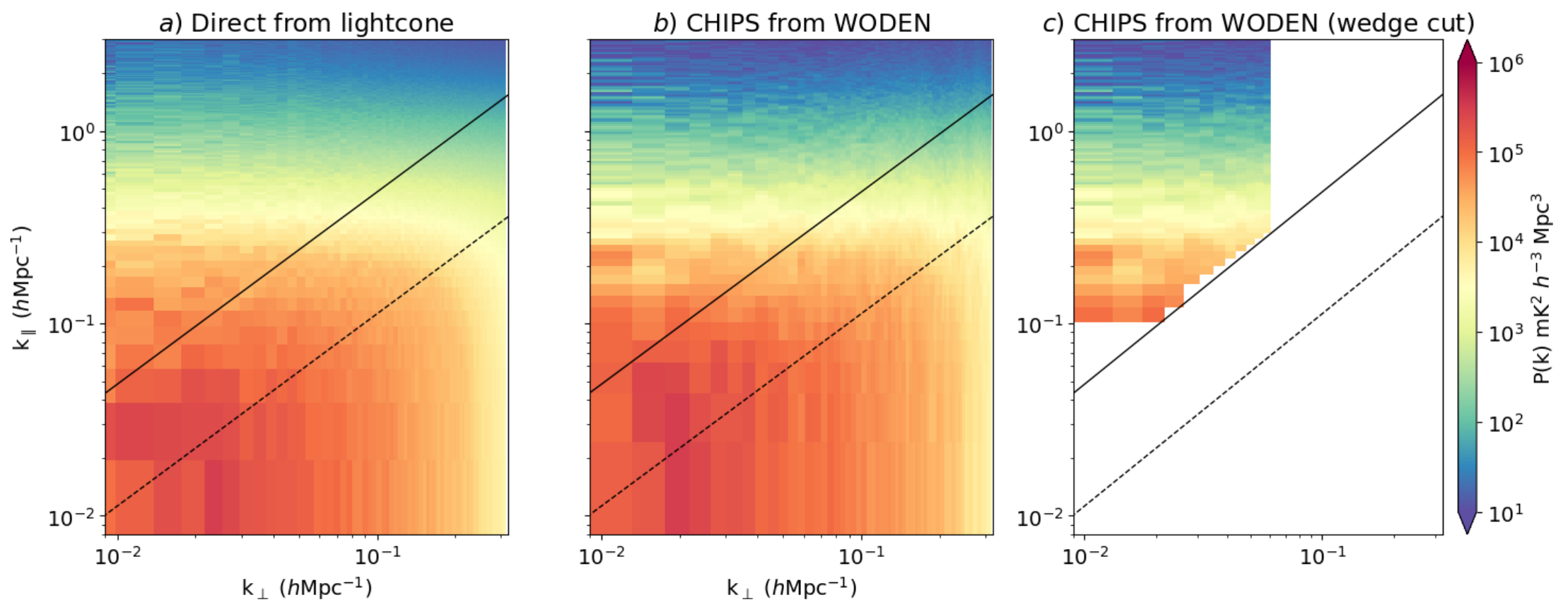}
  \caption{2D PS from the \twocm simulation where: \textit{a)} produced using \OSIRIS directly from the lightcone box from \citet{Greig2022}; \textit{b)} a \CHIPS 2D PS made from integrating over 30 simulated observations; \textit{c)} the same PS from \textit{(b)} but with standard cuts made to remove foreground contamination. The solid black line indicates an estimate of the horizon, and the dashed black line the full-width half-max of the MWA primary beam.}
  \label{fig:2D_comparison_to_lightcone}
\end{figure*}

To test the \twocm model detailed in Section~\ref{sec:21cm_skymodel}, we simulate 30 observations, matching the first two nights detailed in Figure~\ref{fig:obs_selection}, with the same parameters as used for the Gaussian noise simulation (detailed in Section~\ref{sec:gauss_noise}). We leave out simulating the third night purely to save on computational resources. To qualitatively assess the accuracy of the model through \WODEN and into \CHIPS, we compare a \CHIPS 2D PS to one derived directly from the lightcone box, shown in Figure~\ref{fig:2D_comparison_to_lightcone}. We produce the 2D PS directly from the lightcone using the \OSIRIS package\footnote{\url{https://github.com/JaidenCook/OSIRIS}}~\citep{Cook2022}, where no primary beam or instrument sampling was applied, meaning the entire $u,v$ plane was sampled. We see that the \CHIPS PS is broadly consistent with the \OSIRIS PS. Noticeably, it seems there is a potential signal loss at high $k_\parallel$ in the \CHIPS PS (the darker blue region at the top of Figure~\ref{fig:2D_comparison_to_lightcone}{\color{blue}b}). There is also excess power seen along high $k_\perp$, around the dashed black line, which is expected as the instrument baseline sampling causes mode mixing, moving power up from lower to higher $k_\parallel$.

\begin{figure}[h!]
    \centering
    \includegraphics[width=\columnwidth]{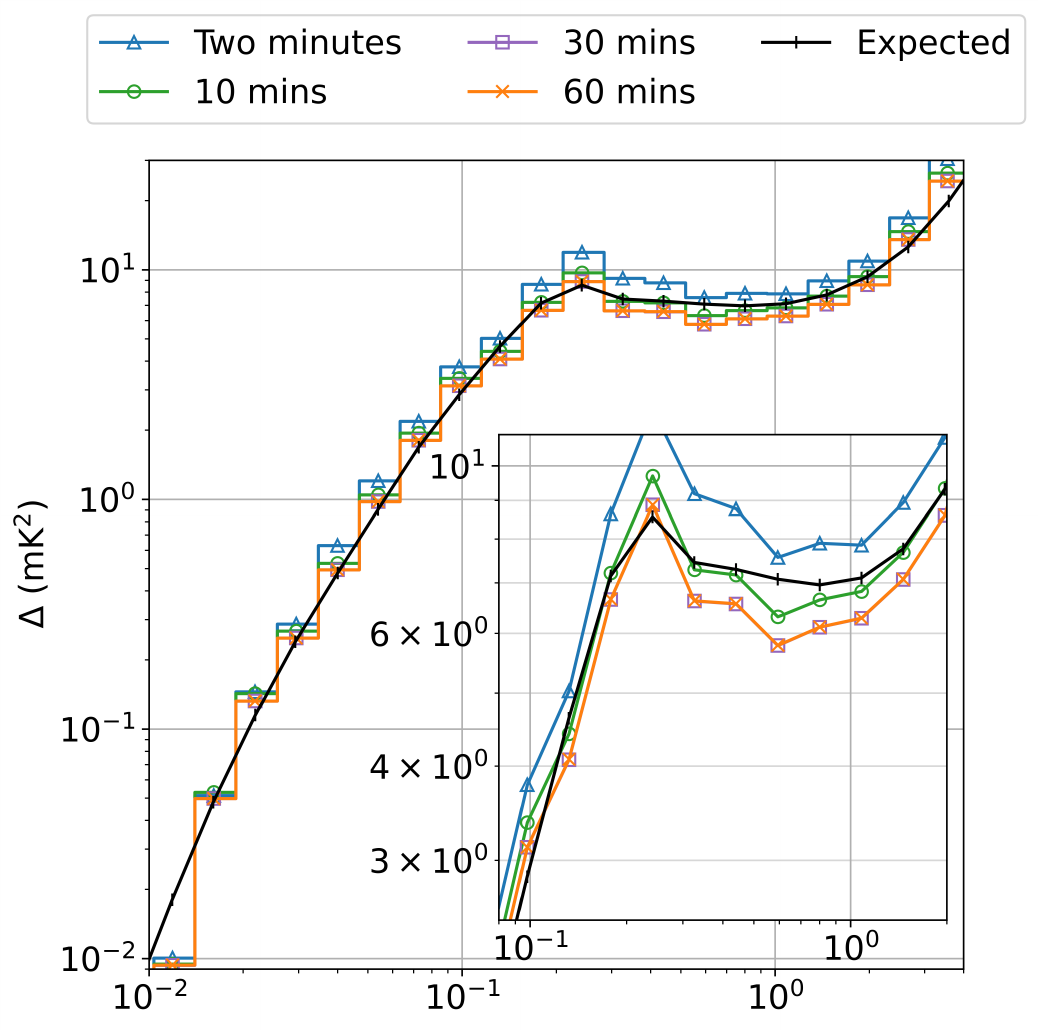}
    \caption{Recovered \twocm signal from various integration lengths of simulated observations, when normalised with a factor 2.5845. The panel bottom right is included to illustrate the normalisation effects between $0.1 < k < 1.0\,h\, \textrm{Mpc}^{-1}$. The two minute data set is for a zenith pointing; the 10 minute data set is for the LST2 subset; all other data sets are an even split between the five pointings and LST subsets as described in Section~\ref{sec:recover_21cm}.}
    \label{fig:gridding_factor_on_21cm}
\end{figure}

To test the gridding normalisation, we apply it to various integrations and produce 1D PS, shown in Figure~\ref{fig:gridding_factor_on_21cm}. We compare them to an expected signal, again derived directly from the \twocm lightcone box. Excellent agreement is shown, and similarly to the Gaussian noise simulation, the gridding density normalisation estimates too much power for short integrations, but renders consistent results for longer integrations with the same mix of LSTs and pointings.

\begin{figure}[h!]
  \centering
  \includegraphics[width=\columnwidth]{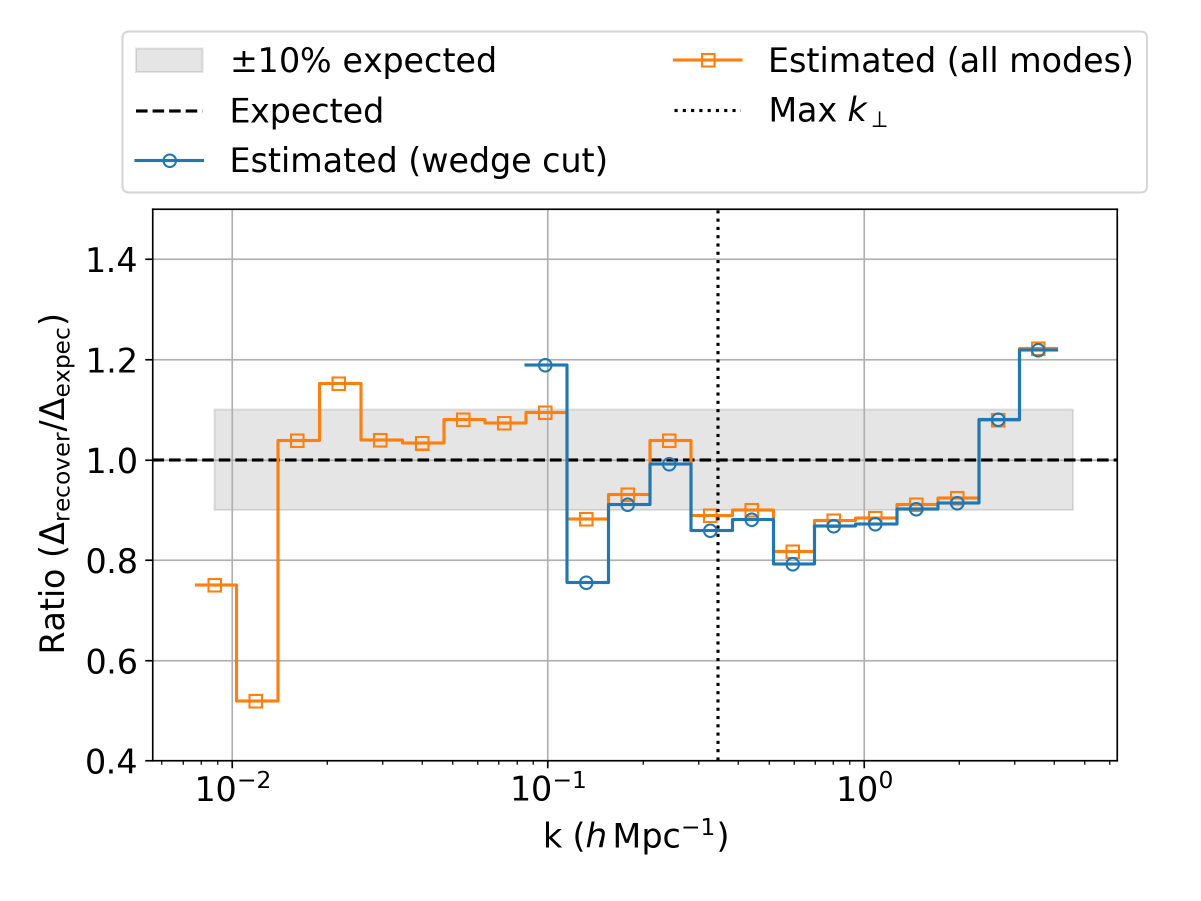}
  \caption{Ratios of the recovered \twocm signal to the expected value, for a 60$\,$min integration. The orange with crosses line shows the ratio when using the factor two gridding density correction, and the blue with circles when using the fitted correction. The vertical dashed line shows the maximum $k_\perp$-mode, meaning any $k$-mode above this was derived purely from the Fourier transform of the visibilities along frequency.}
  \label{fig:final_recover_21cm}
\end{figure}

In Figure~\ref{fig:final_recover_21cm} we show the ratio of the final recovered \twocm signal to the expected value, for a 60$\,$min integration. While good agreement is shown, there does seem to be signal loss around $0.1 < k < 1$, which was not seen for the Gaussian noise simulation. This is likely due to the frequency interpolation of the original lightcone as detailed in Section~\ref{sec:21cm_skymodel}. For reference, Figure~\ref{fig:final_recover_21cm} shows the maximum $k_\perp$ mode that went into the average from 2D to 1D. $k$ modes above this are binned from only high $k_\parallel$ modes, which is where signals with the most spectral structure manifest. As this area is where the signal loss is found, it is likely the frequency interpolation is causing this. As the Gaussian noise sky model has no correlation over frequency, it should not suffer from this interpolation, which is why it doesn't display the same signal loss.

Figure~\ref{fig:final_recover_21cm} also shows the effect of performing the `wedge-cut', where modes normally dominated by foreground sources are removed. While broadly consistent, the signal loss seems to worsen after the cuts. This is likely due to the reduced number of bins going into the average from 2D to 1D. As can be seen from Figure~\ref{fig:2D_comparison_to_lightcone}{\color{blue}c}, the horizon cut (marked by the solid black line) removes more samples for smaller $k_\parallel$ modes. However, lower sampling is not the only possible reason. The manifestation of the primary beam, as well as the exact $u,v$ distribution, could both cause a systematic bias at specific spots in the 2D PS, which could contribute here.

%Note that without the wedge-cut, the two lowest $k$-modes are derived from a single bin from the 2D PS.

\section{Discussion}
\label{sec:discuss}
The Gaussian noise simulation results in Section~\ref{sec:gauss_noise} confirm that gridding visibilities without accounting for their full covariance from overlapping footprints requires a normalisation factor. They reveal however that this normalisation factor depends on an interplay between the array layout, the LST range of the integrated observations, as well as the primary beam pointings, as these all effect the covariance between the visibilities. To truly correct for this effect, one would need to calculate this covariance, and propagate it through the gridding and PS estimation steps~\citep[see][for more information on calculating covariances in $k$-space]{Liu2014}. We instead simply measure and fit for the effect, given we can estimate the expected power from the Gaussian noise simulation. It could be argued that simply taking the median of the ratios where $k_\perp < 0.1\,h\,\textrm{Mpc}^{-1}$ from Figure~\ref{fig:measure_gridding_factor}{\color{blue}d} is too simplistic, given we can fit a functional form as shown in Appendix~\ref{app:fit_gridding_factor}. We choose to err on the side of simplicity however, given the single scaling factor we derive recovers the correct Gaussian-noise power to within 10\% for most $k$-modes. This normalisation factor comes with a set of caveats, which require further investigation to be understood:

\begin{itemize}
  \item We have only tested the MWA phase I layout. Given the covariance between $u,v$ may well be different for different array layouts, we cannot be sure this correction factor would work with the MWA phase II compact layout, for example
  \item Upon initial investigation, 2$\,$s resolution data was seen to require a smaller normalisation factor for the same integration length. It is possible that gridding four more times more data will eventually reach the same plateau as 8$\,$s data, but it might change the level of covariance in the gridded visibilities. Testing this will require four times the computational resources
  \item It is possible that the gridding normalisation will change for a different frequency range, as the rate of change of the $u,v$ points with time will be different. To test this with the method outlined in this paper, another \twocm lightcone box will need to be generated to match the required frequency coverage
  \item We attribute the signal loss seen in Figure~\ref{fig:final_recover_21cm} to the frequency interpolation of the original \twocm lightcone box. However, there could be other sources of signal loss, which include the primary beam shape altering the signal distribution, and a dependence on details of the gridding within a $k_\bot$ bin. To remove frequency interpolation and check this signal loss issue, the original lightcone box would need to be regenerated with constant frequency sampling. This is non-trivial, as it would require \texttt{21CMFAST} simulations with adaptable voxel sizes. Alternatively, one could run the noise sky simulation test using a type of noise with a spectral structure (e.g. Brownian noise). As long as the shape of the noise could be easily predicted in PS space, one could test for signal loss
\end{itemize}

It should be noted that this method depends on not only the Gaussian noise model being correctly scaled, but that our prediction for the power level is correct. However, given the normalisation also yields the expected power level for the \twocm model, with that prediction made directly from the lightcone, this gives us confidence that our prediction and scaling are correct. We further note that we ran the \WODEN simulations in so-called \texttt{float} mode, which sets some of the internal precision to single instead of double precision, with an appreciable speed up~\citep[see][for more details]{Line2022}. We ran a number of comparisons to fully \texttt{double} simulations all the way through to the PS and saw no change in the results. 

It is worth considering that gridding density is not necessarily the only effect that could be causing a loss in signal. The model itself, the \WODEN simulator, and \CHIPS are potentially sources of signal loss. However, given that the Gaussian noise sky model is as simple a model as possible, \WODEN has been heavily unit tested\footnote{See \url{https://woden.readthedocs.io/en/latest/testing/cmake_testing.html} for details on testing}, and we are fitting a \CHIPS-specific normalisation factor, we are confident the normalisation derived here is applicable to real data put through the \pipe.

% {\color{blue} JLBL things that could be mentioned here, or maybe analysis that should be done in this paper?
% \begin{itemize}
%   \item If this correction is true, it means the normalisation is 30\% higher than what we've been using previously. Does that mean our limits should all be 30\% higher? Should we be addressing this in this paper?
%   \item Do this again with a night or two of phase II compact data? Would take a couple of days to get the simulations done.
%   \item As shown in Figure~\ref{fig:pointings}, for the LST1 and LTS3 subsets, the primary beam clips the edge of the EoR model, so part of the model only covers just outside the 10\% beam power level. Could this be a reason we see slightly less power recovered for these subsets, like seen in Figure~\ref{fig:measure_gridding_factor}{\color{blue}b}?
% \end{itemize}
% }

\section{Conclusion}
\label{sec:conclusion}
We have projected a \twocm lightcone into an celestial coordinate-locked sky model for use with the \WODEN simulator. The model is stored in a FITS table format easily adaptable to other simulation packages, and is available from the PASA Datastore. In testing the validity of this model, we generated an equivalent White Gaussian noise model. We used this to test the effects of gridding visibilities using a kernel to produce power spectra, through simulated observations. We confirmed the known effect that power estimated from gridded visibilities is lower where the gridding density is higher. We found the effect to be dependent on a combination of array layout, primary beam pointing, and LST of integrated observations. We also observed that the reduction in power estimated plateaus after integrating more than 30 minutes of data. We found a single scalar normalisation factor of 2.651 was sufficient to recover the expected power to within 10\% for most $k$-modes in a 1D PS. We then generated simulated observations of the \twocm sky model, and verified that \CHIPS was able to recover the expected signal. We found that frequency interpolation of the original lightcone box causes some signal loss at higher $k$, but with the 2.651 normalisation factor, the recovered power is again close to within 10\% of the expect signal. We conclude that this sky model is sufficient to accurately test the \pipe, and can be used in conjunction with foreground models and simulated instrumental effects to test calibration.

\begin{acknowledgement}
This scientific work uses data obtained from \textit{Inyarrimanha Ilgari Bundara} / the Murchison Radio-astronomy Observatory. We acknowledge the Wajarri Yamaji People as the Traditional Owners and native title holders of the Observatory site. Establishment of CSIRO's Murchison Radio-astronomy Observatory is an initiative of the Australian Government, with support from the Government of Western Australia and the Science and Industry Endowment Fund. Support for the operation of the MWA is provided by the Australian Government (NCRIS), under a contract to Curtin University administered by Astronomy Australia Limited. This work was supported by resources provided by the Pawsey Supercomputing Research Centre with funding from the Australian Government and the Government of Western Australia.

During this work we made extension use of the \texttt{kvis}~\citep{Gooch1996} and \texttt{DS9}~\citep{Joye2003} FITS file image viewers.
\end{acknowledgement}

\paragraph{Funding Statement}

This research was supported by the Australian Research Council Centre of Excellence for All Sky Astrophysics in 3 Dimensions (ASTRO 3D), through project number CE170100013. CMT is supported by an ARC Future Fellowship under grant FT180100321.

\paragraph{Data Availability Statement}
The \twocm and White Gaussian noise sky models are available in the PASA Datastore in a FITS table format. The \CHIPS codebase is available upon request to the authors; all other code bases are linked throughout the paper.

%\endnote in some journals will behave like \footnote; and \printendnotes will not output anything. 
\printendnotes

\appendix
\section{Fitting for gridding density correction factor}
\label{app:fit_gridding_factor}
\begin{figure}[h!]
  \centering
  \includegraphics[width=\columnwidth]{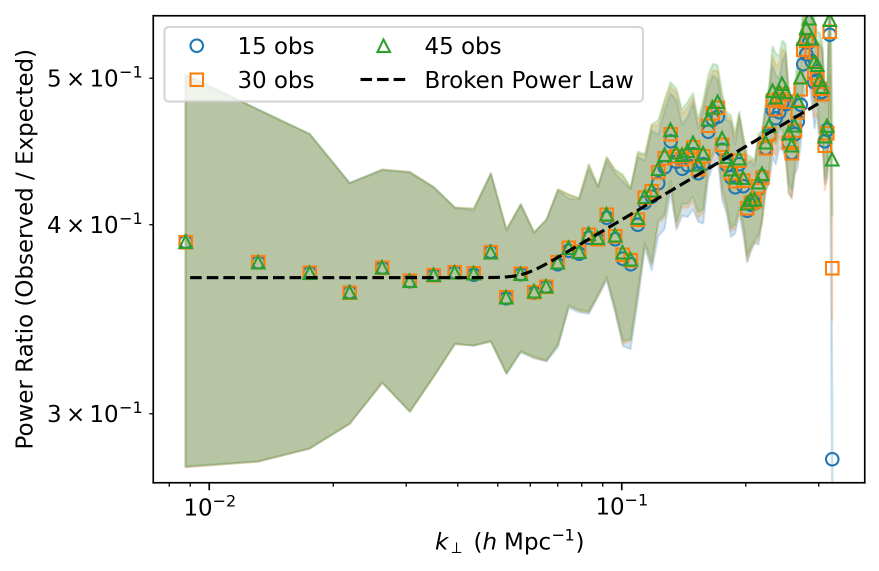}
  \caption{Fitting the median recovered ratio of the noise simulation as a function of $k_\perp$. Circles, squares, and triangles show the median ratio for integrations of 15, 30, and 45 observations respectively. The shaded regions are bound by the median absolute deviation. The dashed black line shows the fitted broken power-law.}
  \label{fig:fit_as_func_kperp}
\end{figure}

As noted in Section~\ref{subsec:grid_dense_correct}, the power estimated from the Gaussian noise simulations is consistent for integrations of 15, 30, and 45 observations (when using the same combination of LSTs and pointings). We can therefore correct for gridding density by fitting the observed ratio of measured to expected power as a function of $k_\perp$. We fit a broken power-law (see Figure~\ref{fig:fit_as_func_kperp}) as defined by\footnote{See \url{https://gist.github.com/cgobat/12595d4e242576d4d84b1b682476317d} for more information on the functional form}
\begin{equation}
  % A*(k_perp_mesh/x_b)**(-a_1) * (0.5*(1+(k_perp_mesh/x_b)**(1/delta)))**((a_1-a_2)*delta)
  A \frac{k_\perp}{k_\mathrm{break}}^{-\alpha_1} \left[0.5\left(1+\left(\frac{k_\perp}{k_\mathrm{break}}\right)^{\frac{1}{\Delta}}\right)\right]^{(\alpha_1-\alpha_2)\Delta},
\end{equation}
where $A$ is an amplitude factor, $k_\mathrm{break}$ is the $k_\perp$ value where the power-law breaks, $\alpha_1$ is the power-law index for $k_\perp < k_\mathrm{break}$, $\alpha_2$ is the power-law index for $k_\perp > k_\mathrm{break}$, and $\Delta$ is the sharpness of the break. We set $\alpha_1 = 0$ as the gridding density plateaus at low $k_\perp$, and set $A = 0.37728$, the median of the ratios where $k_\perp < 0.1\,h\,\textrm{Mpc}^{-1}$ from Figure~\ref{fig:measure_gridding_factor}{\color{blue}d}. We obtain $\alpha_2$ by performing a least squares fit with a single power-law model $C k_{\perp}^{\alpha_2}$, where $C$ is some constant, on ratios where $k_\perp > 0.07\,h\,\textrm{Mpc}^{-1}$. We can then find $k_\mathrm{break}$ by finding where the power-law is equal to the amplitude $A$:
\begin{align}
  A = C k_\mathrm{break}^{\alpha_2}, \\
  k_\mathrm{break} = 10^\frac{\log_{10}(A) - C}{\alpha_2 }.
\end{align}
We fit this function to the median value of the ratios for each $k_\perp$ bin, as we found the values in each bin to have non-zero skew. Taking the median should therefore be a better representation of the data rather than the mean. We summarise the fitted parameters in Table~\ref{tab:fit_params}. To create a normalisation function, we simply take the inverse of this broken power-law.

\begin{table}[h]
\centering
\caption{Parameters for broken power-law fit}
\begin{tabularx}{.6\columnwidth}{>{\setlength\hsize{.37\hsize}} X | >{\setlength\hsize{0.63\hsize}} X}
\hline
\textbf{Parameter} & \textbf{Value} \\
\hline
\hline
$a_1$ & 0.0 \\
$\Delta$ & 0.01 \\
$A$ & 0.377280447218 \\
$a_2$ & -0.161193559625 \\
$k_\mathrm{break}$ & 0.064354113566 \\
\hline
\end{tabularx}
\label{tab:fit_params}
\end{table}

\begin{figure}[h!]
  \centering
  \includegraphics[width=\columnwidth]{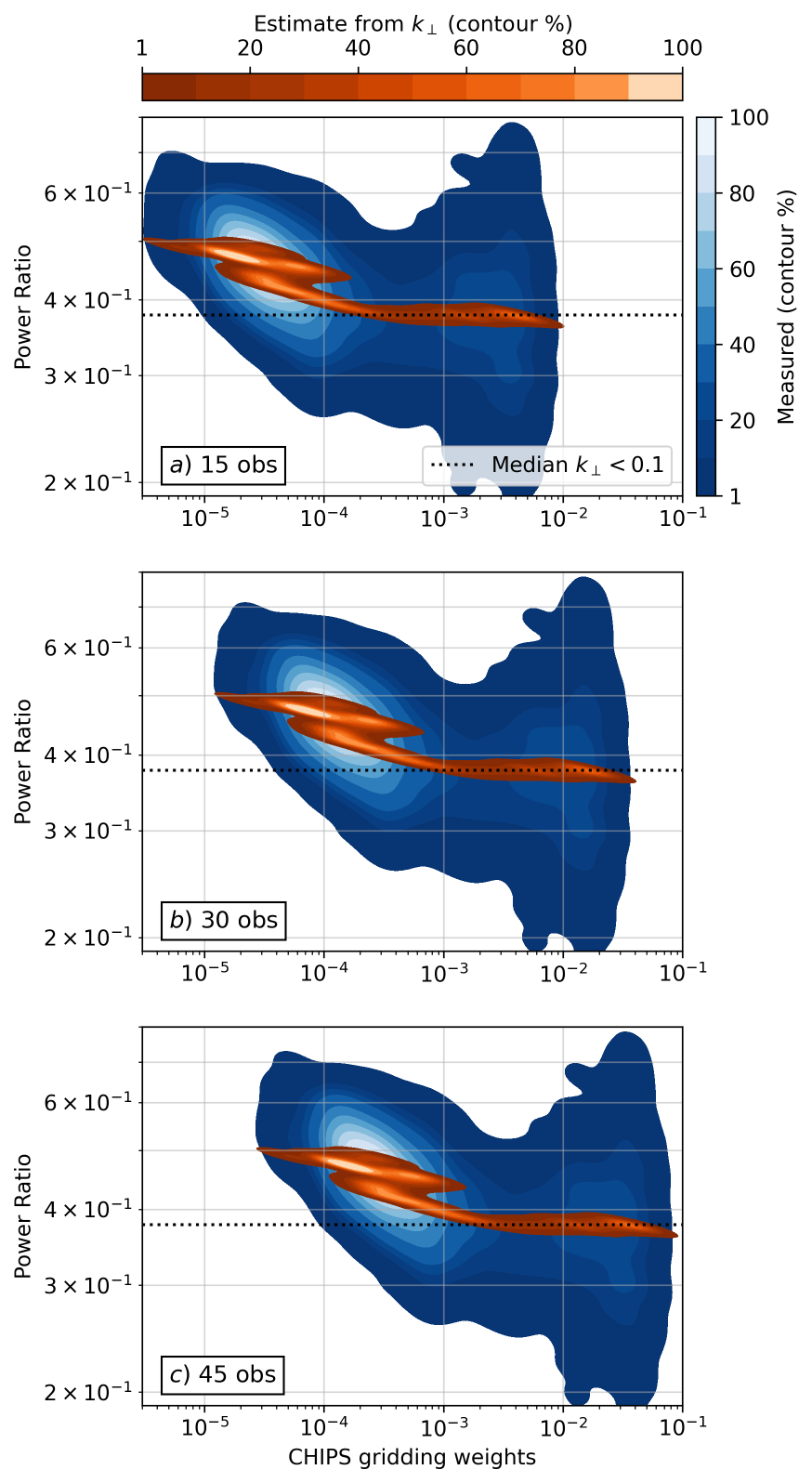}
  \caption{Kernel density estimates of the recovered ratio of the noise simulation as a function of \CHIPS gridding weights (blue), and those predicted by fitted broken power-law (orange) for: \textit{a)} the 15 observation integration of all LSTS; \textit{b)} 30 observations; \textit{c)} 45 observations.}
  \label{fig:fit_alllst_weights}
\end{figure}

\begin{figure}[h!]
  \centering
  \includegraphics[width=\columnwidth]{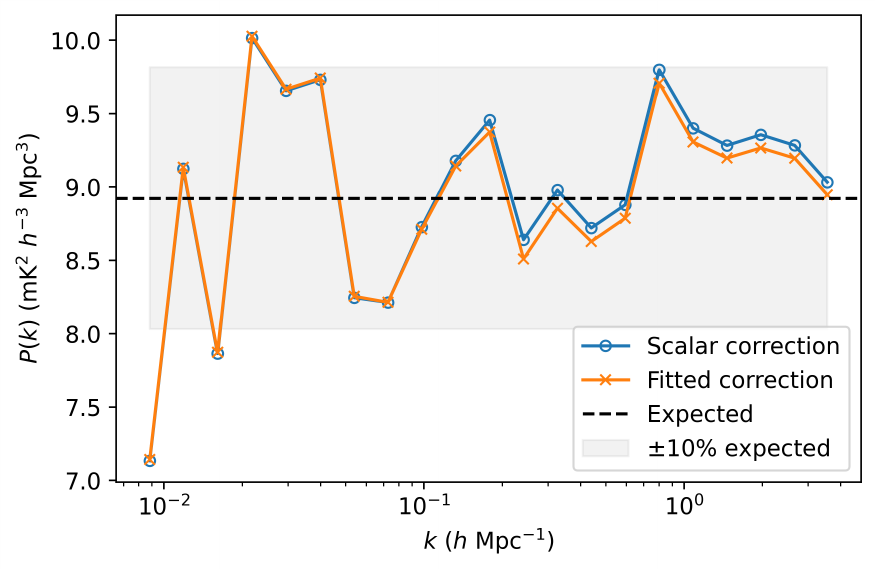}
  \caption{The recovered 1D PS from the Gaussian noise simulation when correcting with the single scalar normalisation factor, and the fitted functional form as shown in Figure~\ref{fig:fit_as_func_kperp}}.
  \label{fig:scalar-vs-fit_recovery}
\end{figure}

We compare how well this fit predicts the observed ratios in Figure~\ref{fig:fit_alllst_weights}. Here we plot the ratios as a function of the gridding weights, rather than $k_\perp$. We use the \texttt{seaborn}~\citep{Waskom2021} \texttt{kdeplot}\footnote{\url{https://seaborn.pydata.org/generated/seaborn.kdeplot.html}} function to generate two dimensional kernel density estimates (KDE) to show the distribution of observed ratios in blue. We also plot the KDE of the ratios predicted by the fitted broken power-law in orange. We see that the fitted function predicts values close to the median value at high gridding weights. As data with a high gridding weight is naturally up-weighted, these data points contribute more to the estimated power in the 1D PS. The net result of this is shown in Figure~\ref{fig:scalar-vs-fit_recovery}, where we compare the recovered 1D PS from the Gaussian noise simulation when correcting with the single scalar normalisation factor, and the fitted functional form. No difference is seen at low $k$, with minimal difference seen at high $k$. We conclude that as the functional form provides little difference to any limits, the scalar normalisation factor is sufficient for correcting for gridding density.

\printbibliography

% \section{Software versions}

% \begin{table}[h]
% \centering
% \renewcommand{\arraystretch}{1.5}
% \caption{Versions of software used}
% \begin{tabularx}{.99\columnwidth}{>{\setlength\hsize{.37\hsize}} X | >{\setlength\hsize{0.63\hsize}} X}
% \hline
% \textbf{Software} & \texttt{git -{}-describe} \\
% \hline
% \hline
% WODEN & FILL THIS \\
% \hline
% CHIPS & FILL THIS \\
% \hline
% \end{tabularx}
% \label{table:software}
% \end{table}

\end{document}